\documentclass[twocolumn,prl,amsmath,amssymb,showpacs]{revtex4}
\usepackage{mathrsfs}
\usepackage{graphicx}
\usepackage{dcolumn}
\usepackage{bm}
\usepackage{slashed}

\begin{document}

\title
{Phase coherent control in electron-argon scattering in a
bichromatic laser field}
\author{Bin Zhou$^{1}$ and Shu-Min Li$^{1,2}$}
\affiliation{ $^1$Department of Modern Physics, University of
Science and Technology of China, P. O. Box 4,  Hefei, Anhui
230026, People's
Republic of China. \\
$^2$Institut f\"ur Theoretische Physik, Univsit\"at Heidelberg,
69120 Heidelberg, Germany.}
\date{\today}

\begin{abstract}
We study the elastic scattering of atomic argon by a electron in the
presence of a bichromatic laser field.
The numerical calculation is done in the first Born approximation
(FBA) for a simple screening electric potential. With the help of
numerical results we explore the dependence of the
differential cross sections (DCS) on the relative phase $\varphi$
between the two components of the radiation field and discuss the
influence of the number of photons exchanged on the phase-dependence
effect. Moreover, we also discuss the numerical results of the DCS
for different scattering angles and impact energies.
\end{abstract}

\pacs{34.80.Qb; 32.80.Wr; 34.50.Rk; 34.80.Bm}

\maketitle

Multiphoton free-free transitions (MFFT) have attracted much
attention since the pioneering papers by Bunkin and Fedorov
\cite{Bunkin} and Kroll and Watson \cite{Kroll}, and a great deal
of work has been devoted to them
\cite{Kruuger,Mittleman,Wallbank,Madsen,Li02}. As the experimental
technology improved, powerful lasers and new kinds of laser fields
have been applied to laser-assisted atomic and molecular
processes.
Due to the significance in dynamic control of the multicolor
lasers, the processes they modify received considerable
attentions. By changing the phase difference of the multicolor
laser field, we can enhance or modify
atomic and molecular processes, which are called phase coherent
control. There has been much work done in phase coherent control
\cite{Ghalim,Zhang,Protopapas,Kami,Varr,Zhang S
T,VarrS,VarrSand,Cionga,Miloevic}, thus the investigation on the
phase coherent control of elastic electron-atom collisions in a
multicolor laser field becomes a very active research domain.
In most of the theoretical work,
the laser radiation is treated as a classical radiation filed with a
single frequency $\omega$, or some narrow band multi-mode
approximation has been employed, yielding better agreement with the
experiments by Weigngarshofer \cite{Weingartshofer}. Describing a
laser beam by a monochromatic classical background field relies on
the argument that in a laser beam the density of radiation quanta is
so large that the depletion of this beam by emitting or absorbing
quanta from it is negligible. If the laser frequency $\omega$ and
intensity $I$ are sufficiently low so that the excitations of atomic
transitions can be neglected, the atomic target can be described by
a short range potential $V(r)$ and the scattering can be treated in
the first Born-approximation, as was done by Bunkin and Fedorov
\cite{Bunkin}. In this letter, we will investigate the relative
phase $\varphi$ dependence of free-free transitions for the
electron-argon scattering in the presence of bichromatic laser
field.

During a laser-assisted electron-impact scattering process $l$
photons may be exchanged with the laser field. In our work, we
consider the
free-free transition in argon atom in the presence of a bichromatic
laser field, accompanied by transfer of $l$ photons in the first
Born approximation ($l>0$ for emission and $l<0$ for absorption).
The laser field is treated classically as a electromagnetic field
\cite{Mittleman M H}
which is a superposition of two components of frequencies $\omega$
and 2$\omega$. The bichromatic laser field is described as
$\boldsymbol{\mathcal {E}}(t)=\boldsymbol{\mathcal
{E}}_0[\sin\omega t+\sin(2\omega t+\varphi)]$, where
$\boldsymbol{\mathcal {E}}_0$ is the electric field amplitude
vector and the relative phase $\varphi$ can be arbitrarily
changed. Atomic units $\hbar=m=e=1$
are used throughout.

\begin{figure}[htbp]
\centering
\includegraphics[width=17cm,bb= 0 0 819 573]{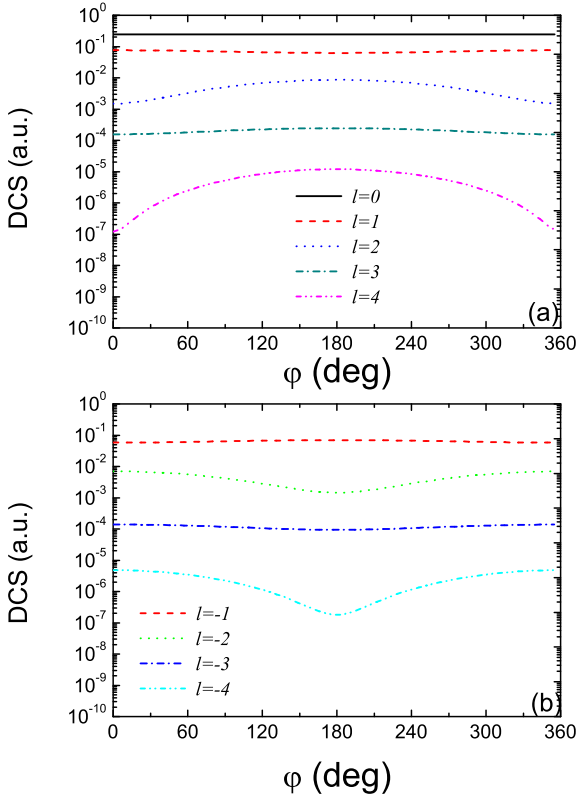}
\caption{\label{fig:epsart}
The $\varphi$-dependence of DCS at a scatteing angle
$\theta=13^\circ$ for electron-argon elastic scattering in a
$CO_2$ laser field ($\hbar\omega=0.117eV$) with $l$ photons
exchanged between the scattering system and the bichromatic laser
field. The impact energies of the incident electron is
$E_i$=9.5eV, and laser amplitud $\mathcal
{E}_0=2.7\times10^5Vcm^{-1}$. (a) Results for emission
($l\geqslant 0$); (b) Results for absorption ($l<0$).}
\end{figure}

\begin{figure}[htbp]
\centering
\includegraphics[width=17cm,bb= 0 0 819 573]{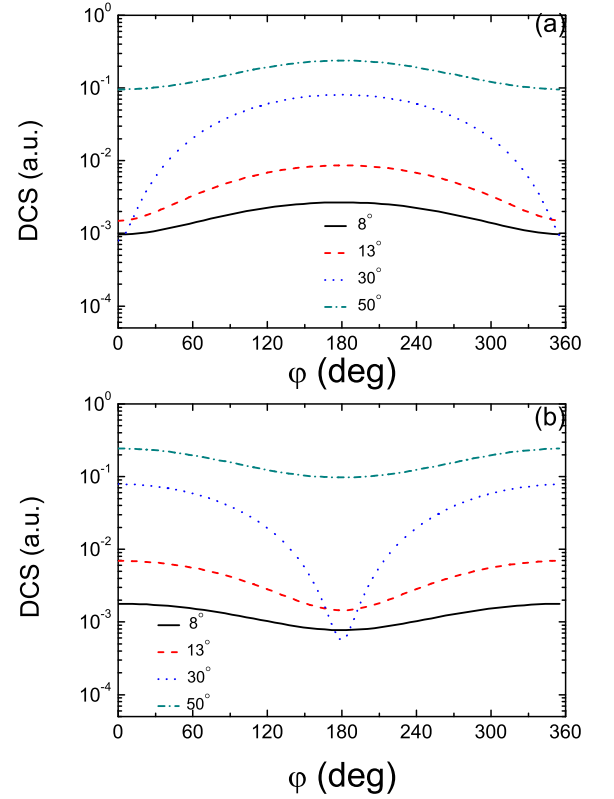}
\caption{\label{fig:epsart}
The $\varphi$-dependence of DCS for $l=\pm2$ at
different scattering angles.
The impact energy and laser parameters
are the same as in Fig.1. (a) Results for emission ($l=+2$); (b)
Results for absorption ($l=-2$).}
\end{figure}

\begin{figure}[htbp]
\centering
\includegraphics[width=17cm,bb= 0 0 821 573]{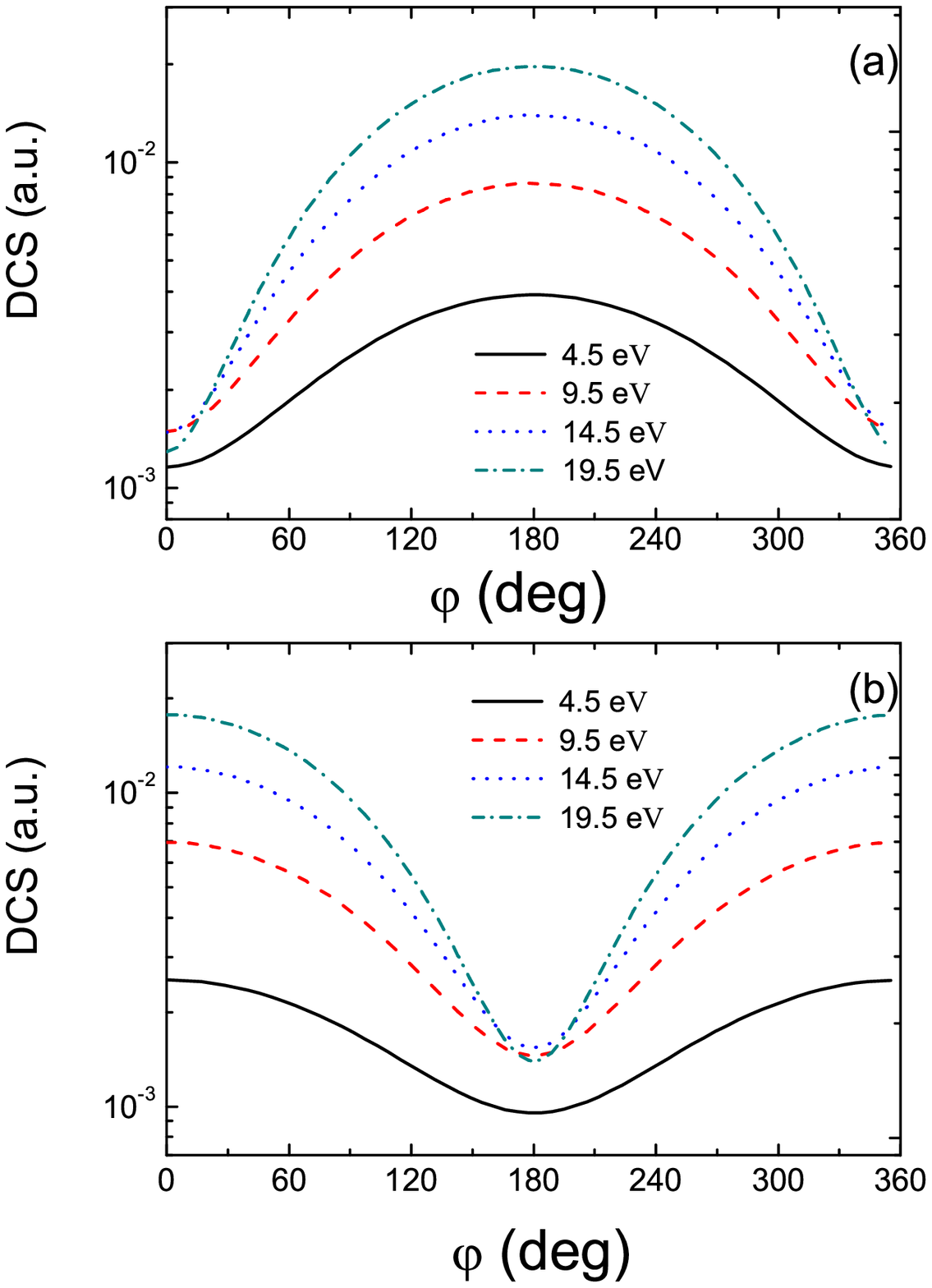}
\caption{\label{fig:epsart}
The $\varphi$-dependence of DCS for $l=\pm2$ at different impact
energies. The scattering angle is $\theta=13^\circ$. The laser
parameters are the same as in Fig.1. (a) Results for emission
($l=+2$); (b) Results for absorption ($l=-2$).}
\end{figure}

In case the laser is much weak compared with the internal field of
an atom (ion), the dressing effect in atom can be neglected, and
the target atom is described by a screening potential
\cite{Salvant,Sun J F,Zhu Z L}
\begin{equation}
V(r)=-\frac{Z}{r}\sum_{i=1}^2A_i\exp(-\alpha_ir), \label{eq1}
\end{equation}
where $r$ denotes the position of the electron with respect to the
nucleus, and $Z$ is the nuclear charge number. For argon,
$A_1=2.1912$, $A_2=-2.8252$, $\alpha_1=5.5470$, $\alpha_2=4.5687$.

The scattering matrix for the laser-assisted free-free transition
in the first Born approximation reads
\begin{equation}
S^{(1)}_{fi}=-i\int_{-\infty}^\infty dt\int d{\bf r}\chi_{{\bf
k}_f}^*({\bf r},t)V(r)\chi_{{\bf k}_i}({\bf r},t), \label{eq2}
\end{equation}
where $\chi_{{\bf k}_f}$ and $\chi_{{\bf k}_i}$ are the initial
and the final states of the electron, described by the Volkov wave
function
\begin{eqnarray}
\chi_{{\bf k}_{f,i}}({\bf r},t)&=&\exp(i{\bf k}_{f,i}\cdot {\bf
r})\exp\left[-iE_{f,i}t-\frac{i}{\omega^2}{\bf
k}_{f,i}\cdot\boldsymbol{\mathcal {E}}_0\sin\omega
t\right]\nonumber\\&~&\times \exp\left[-\frac{i}{4\omega^2}{\bf
k}_{f,i}\cdot\boldsymbol{\mathcal {E}}_0\sin(2\omega
t+\varphi)\right], \label{eq3}
\end{eqnarray}
where ${\bf k}_{i,f}$ are the  wave vectors of electron in the
initial and final states, and $E_{i,f}$ is the corresponding kinetic
energies.

With using the potential of Eq.(\ref{eq1}) and the wave functions
of Eq.(\ref{eq3}), we obtain
\begin{equation}
S^{(1)}_{fi}=-2\pi i\sum_{l}
T_{f,i}^{(1)}(l)\delta(E_f-E_i+l\omega).
 \label{eq4}
\end{equation}
Where $T_{f,i}^{(1)}(l)$ is the transition matrix element,
resolved with respect to multiphoton exchange processes
\begin{equation}
T_{f,i}^{(1)}(l)=B_l(\lambda,\frac{1}{4}\lambda,\varphi)V(k_{f,i}),
\label{eq5}
\end{equation}
in which
\begin{equation}
V(k_{f,i})=\int d{\bf r}e^{-i({\bf k}_f-{\bf k}_i)\cdot {\bf
r}}V(r) \label{eq8}
\end{equation}
is the Fourier transformation of the potential, and
\begin{equation}
B_l(\lambda,\frac{1}{4}\lambda,\varphi)=\sum_{n=-\infty}^\infty
J_{l-2n}(\lambda)J_n(\frac{1}{4}\lambda)\exp(-in\varphi),\label{eq6}
\end{equation}
is the generalized bessel function with $\lambda=({\bf k}_f-{\bf
k}_i)\cdot \boldsymbol{\mathcal {E}}_0/\omega^2$. $J_n(\lambda)$
is the ordinary bessel function.
 The DCS for the net exchange of $l$ photons between the
colliding system and the bichromatic laser field is
\begin{eqnarray}
\frac{d\sigma}{d\Omega}={\frac{1}{(2\pi)^2}}\frac{ k_f}{
k_i}|T_{f,i}^{(1)}(l)|^2.\label{eq9}
\end{eqnarray}

For  numerical calculation, we study the  electron-argon
scattering  under the  geometry of the experiment of
Weingartshofer \cite{Weingartshofer} in which the angle between
the polarization vector $\boldsymbol{\mathcal {E}}_0$ and the
momentum ${\bf p}_i$ of the incident electron is
$\psi_0=38^\circ$, the momentum ${\bf p}_f$ of the scattered
electron is in the plane defined by the polarization vector
$\boldsymbol{\mathcal {E}}_0$ and the momentum ${\bf p}_i$ of the
incident electron. The bichromatic laser
frequencies are respectively $\omega=0.117eV$ and its double
harmonic, with the field amplitude $ \mathcal
{E}_0=2.7\times10^5Vcm^{-1}$, all taken from Ref. \cite{Ghalim}.


In Fig.1, we display cross section dependence on  for each
multiphoton process.
Generally speaking, at the field strength considered, only the
processes with a few photon exchanged have significant
contributions. With the exchanged photon number $l$ increasing,
the DCS decline rapidly. The results for $l=0$ and odd numbers are
less sensitive to the variation of $\varphi$ than $l=$ even
numbers. For the photon emission processes ($l>0$), the DCS at
$\varphi=180^{\circ}$ attain the maximum, while results for photon
absorption attain minimum ($l<0$). All the results are symmetric
about $\varphi=180^\circ$.

Fig.2 shows the $\varphi$-dependence of DCS for $l=\pm2$ at the
scattering angles $\theta=8^\circ$, $13^\circ$, $30^\circ$ and
$50^\circ$. The result for $\theta=30^\circ$ is much more
sensitive to $\varphi$ than the results of other angles. Thus at
this angle, one can more effectively control the dynamics by
varying the phase.


Fig.3 displays the result for $l=\pm2$ at different impact
energies. The tendencies for emission ($l=+2$) and absorption
($l=-2$) are opposite. With the impact energies increasing, the
result becomes more and more sensitive to the variation of phase.

 In summary, we have studied the electron-argon elastic scattering in a bichromatic
laser field with employing a simple potential.  The dependence of
DCS on the relative phase $\varphi$ is investigated at different
scattering angles and impact energies. These features may be used
in the coherent control in the free-free transition.

This work is supported by the National Natural Science Foundation
of China under grant numbers 10874169 and 10674125, and the
National Basic Research Program of China under grant number
2007CB925200. Shu-Min Li would like to thank the financial support
of DFG during his stay in Germany.

\end{document}